\documentclass[12pt]{iopart}

\usepackage{url}
\usepackage{iopams}
\usepackage{amsfonts}
\usepackage{amssymb}
\usepackage{graphicx}
\usepackage{graphics}
\usepackage{hyperref}
\usepackage{tikz}
\usepackage{subfigure}
\usetikzlibrary{shapes.multipart}
\usetikzlibrary{shapes,shadows,arrows}

\begin{document}

\title{Microwave quantum logic spectroscopy and control of molecular ions}

\author{M Shi$^1$, P F Herskind$^1$\footnote{Present address:
Novo Nordisk A/S, Brennum Park, DK-3400 Hiller$\o$d, Denmark.}, 
M Drewsen$^2$ and I L Chuang$^1$}

\address{$^1$ MIT-Harvard Center for Ultracold Atoms, Department of Physics, Massachusetts Institute of Technology, Cambridge, Massachusetts 02139, USA}
\address{$^2$ QUANTOP, Danish National Research Foundation Center of Quantum Optics, Department of Physics, University of Aarhus, DK-8000, Denmark}

\ead{shimolu@mit.edu}

\begin{abstract}

A general method for rotational microwave spectroscopy and control of polar molecular ions via direct microwave addressing is considered. Our method makes use of spatially varying AC Stark shifts, induced by far off-resonant, focused laser beams to achieve an effective coupling between the rotational state of a molecular ion and the electronic state of an atomic ion. In this setting, the atomic ion is used for read-out of the molecular ion state, in a manner analogous to quantum logic spectroscopy based on Raman transitions. In addition to high-precision spectroscopy, this setting allows for rotational ground state cooling, and can be considered as a candidate for the quantum information processing with polar molecular ions. All elements of our proposal can be realized with currently available technology.

\end{abstract}

\maketitle

\section{Introduction}
Precise measurement and control of molecular quantum states are essential to the advancement of a wide range of scientific fields. The development of cooling and control techniques for neutral molecules has, e.g., led to the study of novel quantum phases in the ultra-low temperature regime~\cite{Carr2009}, and both neutral polar molecules~\cite{DeMille2002} and polar molecular ions~\cite{Schuster2011} are being considered as candidate systems in a quantum information processor. Molecular ions are also proposed for quantum chemistry applications~\cite{Willitsch2008} and their spectroscopic data is of great importance in astrophysics~\cite{Canuto1993}. Moreover, in the realm of fundamental science, they are under investigation in tests of the time-variation of physical constants~\cite{Schiller2005,Beloy2011} as well as in the search for the electron electric dipole moment~\cite{Stutz2004,Meyer2006}.

Continued progress in these diverse fields of inquiry is strongly driven by the development of novel spectroscopic techniques on the long lived internal states of molecules. Molecular ion spectroscopy is particularly appealing, for the long storage time of the trapped molecules and their near-perfect isolation from environment can in principle allow experimenters to perform continuous measurements, e.g., to improve the resolution in spectroscopic data~\cite{Koelemeij2007} or to follow system dynamics in real time~\cite{Hoejbjerre2008}. Furthermore, as a key requirement for precision spectroscopy, molecular ions can be translationally cooled to mK temperatures through Coulomb interaction with co-trapped atomic ions~\cite{Bertelsen06, Molhave2000, Roth2006, Chen2011}.

Aside from a long lived state and an effective cooling method to minimize velocity-induced frequency shifts, high-precision molecular ion spectroscopy is currently faced by two challenges: efficient state detection, and reliable state initialization~\cite{Schmidt2005}. Thus far, most methods implemented in the laboratories for quantum state detection have been destructive, and typically based on either photo-dissociation~\cite{Bertelsen06, Hoejbjerre2009, Roth2006, Chen2011, Staanum2010, Schneider2010} or charge transfer~\cite{Tong2010}, which results in loss of the trapped molecule, precluding the experiments from fully exploiting the long storage time of ion traps. Several schemes~\cite{Vogelius2006a, Vogelius2006b, Schmidt2006, Leibfried2012}, as well as experiments~\cite{Staanum2010, Schneider2010, Tong2010} to initialize molecular ions to their internal ground state have been proposed. However, common to the vast majority of these schemes, is that they require dedicated laser system that are specific to the molecular species of interest. It is therefore highly desirable to have in molecular ion spectroscopy, a scheme where a non-destructive state detection and a non-species specific state initialization can both be incorporated.

Several non-destructive methods for molecular state detection have been envisioned, where the state of the molecular ion can be mapped onto that of a cotrapped single atomic ion~\cite{Schmidt2006} or the temperature of an atomic ion crystal~\cite{Clark2010}, which facilitate the read-out. The first scheme, known as quantum logic spectroscopy (QLS)~\cite{Schmidt2005} has recently received renewed interest as a method for molecular ion spectroscopy owing to related works that have proposed its use together with quantum gates driven by frequency comb light~\cite{Leibfried2012,Ding2012} as well as with quantum phase gates~\cite{Mur-Petit2012}. However, as Raman lasers are used to address the molecular ion transitions, the majority of the current QLS implementations are also species specific.

Building on the early work on QLS, we propose a scheme that allows high-precision spectroscopy and ground state initialization of rotational states in molecular ions using microwave fields. We present a microwave pulse sequence for state initialization using heralded state projection~\cite{Vogelius2006b}. The use of microwaves avoids the need for molecule-specific laser systems and can thus be implemented experimentally to work with a broad range of different molecular species without having to change any elements of the experiment apart from the molecular ion source and possibly a, commercially available, microwave source.

One challenge to address the molecular rotational state transitions with microwave fields is that the long wave length implies that the mapping of the rotational state to the ion's electronic state via the motional bus~\cite{Diedrich1989} is in general too weak for any practical implementation. We show that this coupling can be dramatically enhanced via the interaction with a far off-resonant optical field, in a scheme analogous to the methods used for atomic ions in magnetic field gradients~\cite{Mintert2001,Ospelkaus2008}. The presence of the laser field furthermore induces light shifts that effectively provides for individual addressing of closely spaced molecular ions, even when driven by spatially near-isotropic microwave fields. This aspect promotes polar molecular ions as a realistic candidate for a large-scale quantum computing. 

This paper is organized as follows: In Section \ref{sec:QLS} we describe the experimental protocol of QLS on rotational states of molecular ions. In Section~\ref{sec:GroundStateCooling} we outline a scheme for rotational ground state cooling using microwaves and quantum logic. In Section~\ref{sec:LambDicke} we show how to achieve the coherent mapping of the molecular rotational state to motional bus via the interaction with focused, far off-resonant laser beams.  In Section~\ref{sec:QIPpolmol} we discuss implications of our scheme for quantum information processing with polar molecular ions. Finally, the paper is concluded in Section~\ref{sec:Conclusions}.

\section{Quantum Logic Spectroscopy on molecular ions}\label{sec:QLS}

\begin{figure}
\centering
\includegraphics[angle = -90, trim = 10mm 20mm 45mm 15mm, clip, width=0.8\columnwidth]{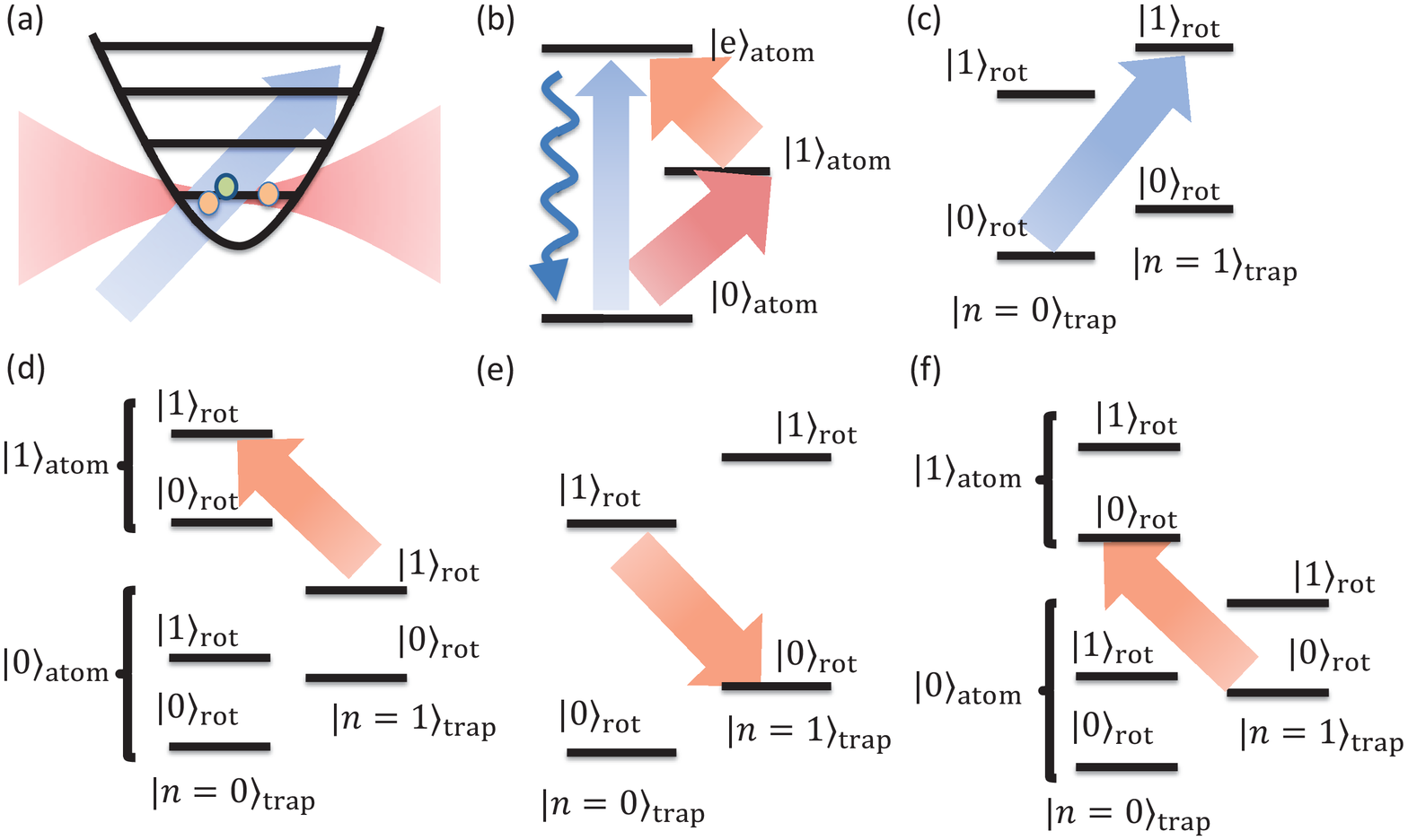}
\caption[justification = justified]{Scheme for QLS quantum logic spectroscopy on molecular ions. (a) Single molecular and single atomic ion co-trapped in a harmonic oscillator with levels $|n\rangle_\mathrm{trap}$. Transitions of the molecular ion are addressed with microwave fields. A tightly focused off-resonance laser field is used to create a sufficiently large effective Lamb-Dicke parameter (Section~\ref{sec:LambDicke}) (b) Quantum logic state of the atomic ion. The short-lived excited state $|e\rangle_\mathrm{atom}$ couples to the lower energy level $|0\rangle_\mathrm{atom}$ and is used for state detection. State $|1\rangle_\mathrm{atom}$ can be pumped to $|e\rangle_\mathrm{atom}$ via optical fields for initialization.  (c) Blue sideband rotational spectroscopy: a $\pi$-pulse on the BSB is used to map the $|J=0\rangle_\mathrm{rot}|0\rangle_\mathrm{trap}$ state to $|J=1\rangle_\mathrm{rot}|1\rangle_\mathrm{trap}$. (d) Motional to atomic state mapping. $|1\rangle_\mathrm{trap}|0\rangle_\mathrm{atom}$ is mapped to $|0\rangle_\mathrm{trap}|1\rangle_\mathrm{atom}$ with a RSB $\pi$-pulse. (e) Red sideband rotational to motional state mapping: a $\pi$-pulse on the RSB is used to map the $|J=1\rangle_\mathrm{rot}|0\rangle_\mathrm{trap}$ state to $|J=0\rangle_\mathrm{rot}|1\rangle_\mathrm{trap}$. (f) Motional to atomic state mapping. Similar to (c), where $|1\rangle_\mathrm{trap}|0\rangle_\mathrm{atom}$ is mapped to $|0\rangle_\mathrm{trap}|1\rangle_\mathrm{atom}$ with a RSB $\pi$-pulse.}
\label{fig:Schematics}
\end{figure}

Here we consider the challenge of doing spectroscopy on the rotational states of a molecular ion using quantum logic spectroscopy. Owing to their rich level structure, molecular ions do not, in general have a cycling transition on which a laser can scatter photons for efficient quantum state detection. QLS has proven to be a particularly useful technique in systems where reading out the quantum state is challenging. One can co-trap a single molecular ion with a single atomic ion in such setting. Via their common motion in the trap, the state of the molecular ion can be mapped onto the atomic ion, which then facilitates efficient state read-out. The specific experimental protocol is detailed in the remainder of this section in four steps, as shown schematically in Figure~ \ref{fig:Schematics} (c)-(f). We start with the assumption that the molecular ion can be prepared in its rotational ground state. Ground state cooling can be achieved via heralded state projection, the details of which are presented in Section~\ref{sec:GroundStateCooling}.

The protocol consists of the following four parts:

(i) \emph{State preparation:} A single atomic ion and a single diatomic molecular ion are confined in the harmonic potential of an ion trap. A schematic is shown in Figure~\ref{fig:Schematics} (a). The transitions of the molecular rotational states can be addressed directly using microwave fields. A sufficiently large effective Lamb-Dicke parameter is created by a spatially varying AC Stark shift due to a tightly focused laser at the molecule (details of this technique are explained in Section~\ref{sec:LambDicke}). Both ions are in their electronic ground state. When in thermal equilibrium with the black body radiation, we assume the molecular ion is in its vibrational ground state owing to the large energy scale of this degree of freedom relative to the black body radiation spectrum, but the rotational wave function is in general, a mixed state to begin with. To initialize the experiment, we cool it to its ground state by heralded projection (Section~\ref{sec:GroundStateCooling}). Here we assume that we are in the $|J = 0\rangle_\mathrm{rot}$ state, or for short $|0\rangle_\mathrm{rot}$, where $J$ is the rotational angular momentum quantum number of the molecule.

We treat the atomic ion as a qubit with states $|0\rangle_\mathrm{atom}$ and $|1\rangle_\mathrm{atom}$, and with a short-lived excited state $|e\rangle_\mathrm{atom}$, which forms a cycling transition together with the $|0\rangle_\mathrm{atom}$ state and thereby allows for efficient discrimination between the qubit states as shown in Figure~\ref{fig:Schematics} (b).
We furthermore assume that the atomic ion has been cooled to near its motional ground state and the translational energies of the atomic and molecular ions have thermalized. After this initial cooling, additional sideband cooling of a common vibrational mode of the atom-molecule system has brought this mode to the ground state. The initial state is thus:
\begin{equation}
\label{eq:initial}
|000\rangle \equiv |0\rangle_\mathrm{atom} |0\rangle_\mathrm{rot}|0\rangle_\mathrm{trap} ,
\end{equation}
where the first, second, and third qubit represents the atomic ion's electronic state, molecular ion's rotational state, and the motional state shared by the atom and the molecule, respectively.

(ii) \emph{Blue sideband rotational spectroscopy:} The process is illustrated in Figure~\ref{fig:Schematics} (c). After the system is prepared in the state given by \Eref{eq:initial}, a microwave pulse is applied on or near the blue sideband (BSB) of the $|0\rangle_\mathrm{rot} \rightarrow |1\rangle_\mathrm{rot}$ transition  of the molecular ion, which in this example is the spectroscopic transition. Since neither the exact frequency nor the strength of the transition is known initially, this will not be a perfect $\pi$-pulse but will transfer the $|0\rangle_\mathrm{rot}$ population to the $|1\rangle_\mathrm{rot}$ state, with some probability $\alpha_1$ close to 1. The net result of this BSB pulse is to effectively entangle the rotational state with the motional state, resulting in:
\begin{equation}
\label{eq:bsb}
\sqrt{\alpha_1}e^{i\theta_1}|011\rangle + \sqrt{1-\alpha_1}|000\rangle,
\end{equation}
where $\theta_1$ is the relative phase between the two states. We are assuming here that the motional sidebands are well-resolved on the $|0\rangle_\mathrm{rot} \leftrightarrow |1\rangle_\mathrm{rot}$ transition and that we are significantly detuned from the carrier transition. The trap frequency is well-known as this can be measured using the atomic ion. 

(iii) \emph{Quantum state mapping to atomic ion:} To allow efficient state detection, we map the motional state onto the atomic ion by applying a RSB $\pi$-pulse on the atomic ion qubit transition. The detuning and Rabi frequency for this transition are known with high precision and we assume for simplicity that this is done with unit fidelity. This is shown in Figure~\ref{fig:Schematics} (d). The resulting state reads:
\begin{eqnarray}
\label{eq:b4Det}
\sqrt{\alpha_1}e^{i\theta_1}|110\rangle + \sqrt{1-\alpha_1}|000\rangle, 
\end{eqnarray}

(iv) \emph{Detection and state re-initialization:} At this point, all the spectroscopic information contained in the transition probability $\alpha_1$ can be revealed by atomic ion state detection. This is facilitated by applying a repump field between $|0\rangle_\mathrm{atom}$ and $|e\rangle_\mathrm{atom}$, and collecting the scattered photons corresponding to the $|e\rangle_\mathrm{atom} \rightarrow |0\rangle_\mathrm{atom}$ relaxation. Depending on whether the atomic ion is in $|0\rangle_\mathrm{atom}$ or $|1\rangle_\mathrm{atom}$, scattering may or may not occur. By repeating the measurement on the atomic ion state for different spectroscopy probe frequencies, we can fully characterize the line shape of the molecule's rotational state transition. 

In order to determine $\alpha_1$ with high precision for each probe frequency, the state in \Eref{eq:b4Det} needs to be regenerated many times to build sufficient detection statistics. An efficient method of re-initializing the wave function after the collapse of \Eref{eq:b4Det} upon each detection is therefore necessary. In our protocol,  this is carried out using heralded projection. We detail the process below.

Depending on the measurement outcome, two different protocols are carried out: In the simplest case, the measurement outcome is $|0\rangle_\mathrm{atom}$ and the system has been projected onto the $|000\rangle$ state. In this scenario, the system has been reinitialized and the spectroscopy protocol can start again. If, on the other hand, $|1\rangle_\mathrm{atom}$ is detected, the system is projected onto the $|110\rangle$ state. Reinitialization to the $|000\rangle$ state can proceed by first removing the qubit excitation via a repump field from state $|1\rangle_\mathrm{atom}$ to $|e\rangle_\mathrm{atom}$, which will relax to $|0\rangle_\mathrm{atom}$, leaving the overall wave function in the rotational excited state $|010\rangle$. Possible excitation of the motional state during this process can be removed through a brief sequence of sideband cooling on the atomic ion. To remove the rotational state excitation, similar to step (ii) and (iii), one can transfer it to the atomic ion via the motional bus. More specifically, one can apply a red sideband $\pi$-pulse on the molecule's rotational state transition as illustrated in Figure~\ref{fig:Schematics} (e), resulting in 
\begin{equation}
\label{eq:rsb_cool1}
\sqrt{\alpha_1}e^{i\theta_1}|001\rangle+\sqrt{1 - \alpha_1}|010\rangle.
\end{equation}
Note that the RSB Rabi frequency is lower than that of the BSB (c.f.~\Eref{eq:bsb}) by a factor of $\sqrt{2}$. We therefore adjust the pulse length accordingly to maintain the same probability $\alpha_1$ for the $\pi$-pulse transition. 
Now, similar to \Eref{eq:b4Det}, making a RSB transition on the atomic state (see Figure~\ref{fig:Schematics} (f)) changes the wave function into
\begin{equation}
\label{eq:rsb_cool2}
\sqrt{\alpha_1}e^{i\theta_1}|100\rangle+\sqrt{1 - \alpha_1}|010\rangle.
\end{equation}
Upon the detection on the atomic state of \Eref{eq:rsb_cool2}, the system is either projected to $|100\rangle$, which can be brought down to $|000\rangle$ by optical pumping, or $|010\rangle$, which can be cooled again by going through the same pulse sequences in \Eref{eq:rsb_cool1} and (\ref{eq:rsb_cool2}).

\tikzstyle{line} = [draw, -stealth, thick]
\tikzstyle{block0} = [draw, rectangle, fill = green!10, text width=20em, text centered, node distance = 4em] 
\tikzstyle{block1} = [draw, rectangle, fill = blue!10, text width=20em, text centered, node distance = 4em] 
\tikzstyle{block2} = [draw, rectangle, fill = blue!10, text width=4em, text centered, node distance = 4em] 
\tikzstyle{block3} = [draw, rectangle, fill = blue!10, text width=13em, text centered, node distance = 4em] 
\tikzstyle{block0_i} = [draw, rectangle, fill = green!10, text width=29em, text centered, node distance = 4em] 
\tikzstyle{block2_i} = [draw, rectangle, fill = green!10, text width=10em, text centered, node distance = 4em] 
\tikzstyle{block3_i} = [draw, rectangle, fill = green!10, text width=8em, text centered, node distance = 4em] 
\tikzstyle{text1} = [rectangle, fill = green!0, text width=16em, text centered, node distance = 4em] 

\begin{figure}[h!]
\centering
\resizebox{0.75\columnwidth}{1.25\columnwidth}{
\begin{tikzpicture}

\node(statePrep1)[block0_i]{$|0\rangle\langle 0|_\mathrm{atom} \otimes  \left[p_i(0)|0\rangle\langle 0|_\mathrm{rot}+\sum_{J=i+1}^\infty p_i(J)|J\rangle\langle J|_\mathrm{rot} \right] \otimes |0\rangle\langle 0|_\mathrm{trap}$};

\node(text)[text1, above of = statePrep1, yshift = -2.0em]{starting from thermal state, $i = 0$};

\node(statePrep2)[block0_i, below of=statePrep1, yshift=-2.5em]{
$\alpha_{i+1} p_i(i+1)|0\rangle\langle0|_\mathrm{atom} \otimes|i\rangle\langle i|_\mathrm{rot}\otimes |1\rangle\langle1|_\mathrm{trap} $\\
$+(1-\alpha_{i+1}) p_i(i+1) |0\rangle\langle0|_\mathrm{atom}\otimes|i+1\rangle\langle i+1|_\mathrm{rot} \otimes|0\rangle\langle0|_\mathrm{trap} $\\
$+\sqrt{\alpha_{i+1}(1-\alpha_{i+1})}e^{i\theta_{i+1}}p_i(1)|0\rangle\langle0|_\mathrm{atom} \otimes |i+1\rangle\langle i|_\mathrm{rot}\otimes |1\rangle\langle0|_\mathrm{trap}$\\
$+\sqrt{\alpha_{i+1}(1-\alpha_{i+1})}e^{-i\theta_{i+1}}p_i(0)|0\rangle\langle0|_\mathrm{atom} \otimes |i\rangle\langle i+1|_\mathrm{rot} \otimes|0\rangle\langle1|_\mathrm{trap}$\\
$+ |0\rangle\langle0|_\mathrm{atom} \otimes \left(p_i(0)|0\rangle\langle 0|_\mathrm{rot} + \sum_{J = i+2}^\infty p_i(J)|J\rangle\langle J|_\mathrm{rot} \right) \otimes |0\rangle\langle0|_\mathrm{trap}$};

\node(statePrep3)[block0_i, below of=statePrep2, yshift=-3.5em]{
$\alpha_{i+1} p_i(i+1)|1\rangle\langle1|_\mathrm{atom}\otimes |i\rangle\langle i|_\mathrm{rot}\otimes |0\rangle\langle0|_\mathrm{trap} $\\
$+(1-\alpha_{i+1}) p_i(i+1) |0\rangle\langle0|_\mathrm{atom}\otimes|i+1\rangle\langle i+1|_\mathrm{rot}\otimes |0\rangle\langle0|_\mathrm{trap} $\\
$+ |0\rangle\langle0|_\mathrm{atom} \otimes \left(p_i(0)|0\rangle\langle 0|_\mathrm{rot} +\sum_{J=i+2}^\infty p_i(J)|J\rangle\langle J|_\mathrm{rot} \right)\otimes|0\rangle\langle0|_\mathrm{trap}$};

\node(ion0_i)[block3_i, below of=statePrep3, yshift=-1.0em, xshift = -6.0em]{new mixed state};

\node(ion1_i)[block2_i, below of=statePrep3, yshift=-1.7em, xshift=5.0em]{$|1,i,0\rangle $\\$ \equiv |1\rangle_\mathrm{atom} |i\rangle_\mathrm{rot}|0\rangle_\mathrm{trap}$};

\node(repump_i)[block3_i, below of=ion1_i, yshift=0.5em, xshift=-5.0em]{$|0,i,0\rangle$};

\node(ion1_icoolm)[block0, below of=repump_i, yshift=0.5em, xshift=-0em]{$\sqrt{\alpha_i}e^{i\theta_i}|0,i-1,1\rangle + \sqrt{1 - \alpha_i}|0,i,0\rangle$};

\node(ion1_icoola)[block0, below of=ion1_icoolm, yshift=0.5em, xshift=-0em]{$\sqrt{\alpha_i}e^{i\theta_i}|1,i-1,0\rangle + \sqrt{1 - \alpha_i}|0,i,0\rangle$};

\node(ion1_icool0)[block3_i, below of=ion1_icoola, yshift=0.5em, xshift=-6em]{$|0,i,0\rangle$};

\node(ion1_icool1)[block3_i, below of=ion1_icoola, yshift=0.5em, xshift=6em]{$|1,i-1,0\rangle$};

\node(ion1_icool1repump)[block3_i, below of=ion1_icool1, yshift=0.5em, xshift=-6em]{$|0,i-1,0\rangle$};

\node(initial)[block1, below of=ion1_icool1repump, yshift=0.5em, xshift = -0em]{$|000\rangle \equiv |0\rangle_\mathrm{atom} |0\rangle_\mathrm{rot}|0\rangle_\mathrm{trap}$};

\node(BSBr2m)[block1, below of=initial, yshift=0.5em]{$\sqrt{\alpha_1}e^{i\theta_1}|011\rangle + \sqrt{1-\alpha_1}|000\rangle$};

\node(RSBm2q)[block1, below of=BSBr2m, yshift=0.5em]{$\sqrt{\alpha_1}e^{i\theta_1}|110\rangle + \sqrt{1-\alpha_1}|000\rangle$};

\node(ion0)[block2, below of=RSBm2q, yshift=0.5em, xshift=-6em]{$|000\rangle$};

\node(ion1)[block2, below of=RSBm2q, yshift=0.5em, xshift = 6em]{$|110\rangle$};

\path [line] (ion0_i.west) --(-16em,-19.0em) |- node[yshift=-9em, xshift= -1em, rotate=90] {probe a higher rotational state, $i = i + 1$}(statePrep1.west);

\path [line] (ion1_icool0.west) --(-16em,-33.7em) |- node[yshift=-5em, xshift= -1em, rotate=90] {repeat projection cooling}(repump_i.west);

\path [line] (statePrep1) ->node[ below of= statePrep1, yshift=2.5em, xshift =0.5em] {RSB rotational $\quad$ to motional state} (statePrep2);

\path [line] (statePrep2) ->node[ below of= statePrep1, yshift=2.4em, xshift =0em] {RSB motional $\quad$ to qubit state} (statePrep3);

\path [line] (statePrep3) -- node[yshift=-0.1em, xshift=-4.3em] {measured $|0\rangle_\mathrm{atom}$}(ion0_i);

\path [line] (statePrep3) -- node[yshift=-0.2em, xshift=4.5em] {measured $ |1\rangle_\mathrm{atom}$}(ion1_i);

\path [line] (ion1_i) -- node[yshift=-0.0em, xshift=-3.0em] {repump}(repump_i);

\path [line] (repump_i) -- node[yshift=-0em, xshift=0.5em] {RSB rotational $\quad$ to motional state}(ion1_icoolm);

\path [line] (ion1_icoolm) -- node[yshift=-0em, xshift=0em] {RSB motional $\quad$ to qubit state}(ion1_icoola);

\path [line] (ion1_icoola) -- node[yshift=-0em, xshift=-4.5em] {measured $|0\rangle_\mathrm{atom}$}(ion1_icool0);

\path [line] (ion1_icoola) -- node[yshift=-0em, xshift=4.2em] {measured $|1\rangle_\mathrm{atom}$}(ion1_icool1);

\path [line] (ion1_icool1) -- node[yshift=-0.2em, xshift=-2.5em] {repump}(ion1_icool1repump);

\path [line] (ion1_icool1repump) -- node[yshift=-0.2em, xshift=-0.5em] {repeat projection cooling $\quad$ down  rotational states}(initial);

\path [line] (initial) -- node[yshift=-0.1em, xshift=-0.5em] {BSB rotational $\quad$ spectroscopy}(BSBr2m);

\path [line] (BSBr2m) -- node[yshift=-0.1em, xshift=-0.0em] {RSB motional $\quad$ to qubit state}(RSBm2q);

\path [line] (RSBm2q) -- node[yshift=0em, xshift=5em, text width = 8em] {measured $|1\rangle_\mathrm{atom}$}(ion1);

\path [line] (RSBm2q) -- node[yshift=-0.1em, xshift=-5em] {measured $|0\rangle_\mathrm{atom}$}(ion0);

\path [line] (ion0.west) -- (-16em,-51.2em) |- node[xshift=-0.6em, yshift=-10em, rotate = 90] {}(initial.west);

\path [line] (ion1.east) -- (14.5em,-51.2em) |- node[xshift=0.8em, yshift=-20em, rotate = 270] {repump, $i = 1$}(repump_i.east);

\end{tikzpicture}
}  
\caption{Summary of the molecular ion rotation spectroscopy procedure. Green: initial ground state cooling via projection measurement for a single atomic and molecular ion pair in its molecular rotational thermal equilibrium. Purple: molecular ion rotation spectroscopy process using direct microwave addressing. State re-initialization is achieved using projection measurement.}
\label{fig:summary}
\end{figure}
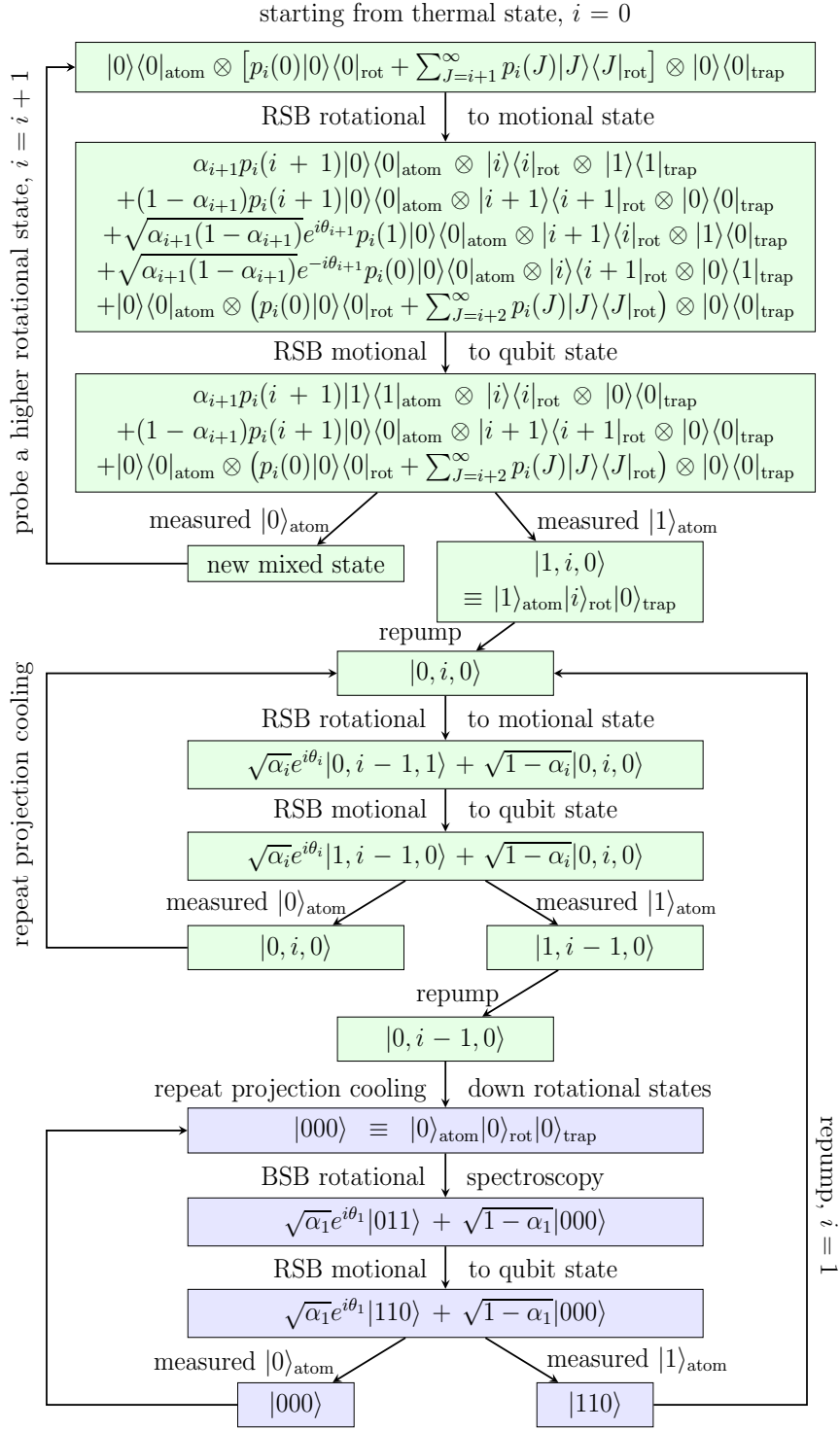

\section{Rotational ground state cooling}\label{sec:GroundStateCooling}
Although ground state preparation of the motional state of molecular ions has proven highly effective through e.g.~sympathetic cooling, rotational ground state cooling still remains a challenge.  Several groups have recently achieved ground rotational state cooling on large molecular ensembles by either direct laser cooling~\cite{Staanum2010,Schneider2010} or by state selective loading of the ion trap~\cite{Tong2010}. Schemes for cooling single molecular rotational states have also been proposed~\cite{Vogelius2006a,Vogelius2006b,Leibfried2012} based on laser driven two photon Raman transitions. So far in our spectroscopy protocol, we have assumed that the trapped ion-molecule system can be initialized to its ground state $|0\rangle_\mathrm{atom} |0\rangle_\mathrm{rot}|0\rangle_\mathrm{trap} $ as in \Eref{eq:initial}. In this section, we describe a scheme where rotational ground state cooling can be achieved using a microwave-based heralded projection scheme. The techniques follow similarly to the state initialization process introduced in Section~\ref{sec:QLS}. 

The lower frequency associated with the molecule's rotation will generally mean that, when in thermal equilibrium with the black body radiation, the molecule is not in its rotational ground state but rather in some distribution, $\sum_{J=0}^\infty p_0(J)|J\rangle\langle J|_\mathrm{rot}$, so the density matrix of the ion-molecule system follows
\begin{equation}
\label{eq:statePrep}
|0\rangle\langle 0|_\mathrm{atom} \otimes  \left[\sum_{J=0}^\infty p_0(J)|J\rangle\langle J|_\mathrm{rot} \right] \otimes |0\rangle\langle 0|_\mathrm{trap},
\end{equation}
where $p_0(J)$ is the probability of finding the molecule in the $|J\rangle_\mathrm{rot}$ state at thermal equilibrium. 
Similar to the techniques presented in the QLS protocol, cooling to ground state can be done by entangling the rotational state with the atomic ion's qubit state, and projecting the former down to its ground state upon state detection of the latter. 
We perform the molecule-to-atom state mapping via the motional bus directly, as outlined in Section~\ref{sec:QLS} (iv). The density matrix in \Eref{eq:statePrep} then turns into:
\begin{eqnarray}
&&\alpha_1 p_0(1) |1\rangle\langle1|_\mathrm{atom} \otimes|0\rangle\langle0|_\mathrm{rot} \otimes|0\rangle\langle0|_\mathrm{trap} \nonumber \\
&&+(1-\alpha_1) p_0(1) |0\rangle\langle0|_\mathrm{atom}\otimes|1\rangle\langle1|_\mathrm{rot}\otimes |0\rangle\langle0|_\mathrm{trap} \nonumber \\   
&&+|0\rangle\langle0|_\mathrm{atom}\otimes \left[\sum_{J\neq1}^\infty p_0(J)|J\rangle\langle J|_\mathrm{rot} \right]\otimes|0\rangle\langle0|_\mathrm{trap}
\label{eq:b4cooling}
\end{eqnarray}
where again, $\alpha_1$ is the probability of transitioning between $|0\rangle_\mathrm{rot}$ and $|1\rangle_\mathrm{rot}$ at the spectroscopy probe frequency, defined earlier in \Eref{eq:bsb}. Now we see that the rotational state is projected to its ground state $|0\rangle_\mathrm{rot}$, upon detection of the atomic ion in $|1\rangle_\mathrm{atom}$. Such detection has been demonstrated with infidelities below $10^{-4}$~\cite{Myerson2008}, which makes this a very robust cooling scheme. The probability of detecting $|1\rangle_\mathrm{atom}$ after this first cooling attempt is 
\begin{equation}
\label{eq:pcool_1}
\mathcal{P}_1^{cool} = \alpha_1 p_0(1), 
\end{equation}
which converges to the population of $|0\rangle_\mathrm{rot}$, $p_0(1)$  when $\alpha_1$ is close to unity at resonance.  For a thermal distribution at temperature $T$, 
\begin{equation}\label{eq:thermEQ}
p_0(J) = \hbar B\beta e^{-\hbar B\beta J(J+1)},
\end{equation}
where $B$ is the rotational constant, and $\beta = 1/k_BT$. For a typical diatomic molecular ion, such as e.g.~SrCl$^+$ of B~=~6.5~GHz, trapped at a cryogenic temperature with T~=~5~K, \Eref{eq:thermEQ} implies $p_0(1) = 0.17$. This means that most times it is more likely to detect $|0\rangle_\mathrm{atom}$, and thus projecting the molecular rotational state to yet another mixed state
\begin{equation}
|0\rangle\langle 0|_\mathrm{atom} \otimes  \left[\sum_{J=0}^\infty p_1(J)|J\rangle\langle J|_\mathrm{rot} \right] \otimes |0\rangle\langle 0|_\mathrm{trap},
\end{equation} 
with $p_1(1) \propto (1-\alpha_1)p_0(1)\rightarrow 0$, when the $|1\rangle_\mathrm{rot} \rightarrow |0\rangle_\mathrm{rot}$ RSB transition is close to resonance. However, following the same pulse sequence, one can proceed to probe the next most populated state, namely $|2\rangle_\mathrm{rot}$, with the addressing microwave frequency tuned close to the $|2\rangle_\mathrm{rot} \rightarrow |1\rangle_\mathrm{rot}$ resonance.  Similarly, the resultant state follows 
\begin{eqnarray}
&&\alpha_2 p_1(2) |1\rangle\langle1|_\mathrm{atom} |1\rangle\langle1|_\mathrm{rot} |0\rangle\langle0|_\mathrm{trap} \nonumber \\
&&+(1-\alpha_2) p_1(2) |0\rangle\langle0|_\mathrm{atom}|2\rangle\langle2|_\mathrm{rot} |0\rangle\langle0|_\mathrm{trap} \nonumber \\   
&&+|0\rangle\langle0|_\mathrm{atom} \left[\sum_{J\neq1, 2}^\infty p_1(J)|J\rangle\langle J|_\mathrm{rot} \right]|0\rangle\langle0|_\mathrm{trap},
\end{eqnarray}
where $\alpha_2$ is the transition probability between $|1\rangle_\mathrm{rot}$ and $|2\rangle_\mathrm{rot}$. Due to the state renomalization after removing the population in 
$|1\rangle_\mathrm{rot}$, the probability of detecting $|1\rangle_\mathrm{atom}$ at this second cooling attempt follows 
\begin{equation}
\mathcal{P}_2^{cool} = \alpha_2 p_1(2) \rightarrow p_1(2) > p_0(2)  = 0.22.
\end{equation}
Detection of $|1\rangle_\mathrm{atom}$ will then project the state to $|0\rangle_\mathrm{atom}|1\rangle_\mathrm{rot}|0\rangle_\mathrm{trap}$. Because this is a pure state, it can be brought down to the ground state by repeating the projection cooling between $|1\rangle_\mathrm{rot}$ and $|0\rangle_\mathrm{rot}$. In the case $|0\rangle_\mathrm{atom}$ is detected, the state returns to the same pure state $|0\rangle_\mathrm{atom}|1\rangle_\mathrm{rot}|0\rangle_\mathrm{trap}$ again, and one can just repeat the cooling pulse sequence until the system is projected to the ground state.
Therefore, to cool the thermal state in \Eref{eq:statePrep}, one can probe the rotational transition of $|1\rangle_\mathrm{rot} \rightarrow |0\rangle_\mathrm{rot}$, $|2\rangle_\mathrm{rot} \rightarrow |1\rangle_\mathrm{rot}$, etc., until the system is projected to a pure state $|0\rangle_\mathrm{atom}|i\rangle_\mathrm{rot}|0\rangle_\mathrm{trap}$, which can then be projected down to the ground state following the ladder downwards.

The initial cooling scheme, as well as the spectroscopy protocol in Section~\ref{sec:QLS} are summarized in Figure~\ref{fig:summary}.

The idea of cooling the molecular rotation via entanglement with an atomic ion was also considered in \cite{Vogelius2006b} and \cite{Leibfried2012}. The main difference between these two earlier works and our proposal is that, while the earlier proposals were based on laser-driven two-photon Raman transitions, here we consider  driving the rotational states directly with microwave fields. This has a number of advantages, including the coupling of rotational states with different parities, which can lead to an efficient way of depopulating the $|J=1\rangle$ state, normally prevented by Raman transitions involving only $^1\Sigma$ electronic potentials. More importantly, Raman schemes require technically demanding dedicated lasers that, in the case of more involved molecular level structure, including hyperfine and magnetic substates, amplify the experimental complexity through the need for frequency tuning and polarization control. Precise control over frequency is one of the main advantages of using microwaves and, given that power requirements are not too severe, generating multiple, finely-tunable frequency components does not incur much significant experimental overhead. 

As an example, we consider again, the case of $\mathrm{SrCl^+}$ trapped at a cryogenic temperature of T~=~5~K. Figure~\ref{fig:population} shows the evolution of the molecular rotational state after measuring $|0\rangle_\mathrm{atom}$, projecting the system onto a new mixed state. The states above $|7\rangle_\mathrm{rot}$ contribute $<0.006$ of the population, and hence this determines the total failure rate. The highest required microwave frequency, i.e. for addressing the $|7\rangle_\mathrm{rot} \rightarrow|6\rangle_\mathrm{rot}$ is 104~GHz, which can be achieved using a commercial tunable microwave source, e.g.~ Agilent E8257D PSG, and a frequency doubler.

\begin{figure}[h!]
\centering
\label{fig:population}
\includegraphics[trim = 15mm 6mm 15mm 4mm, clip, width=0.8\columnwidth]{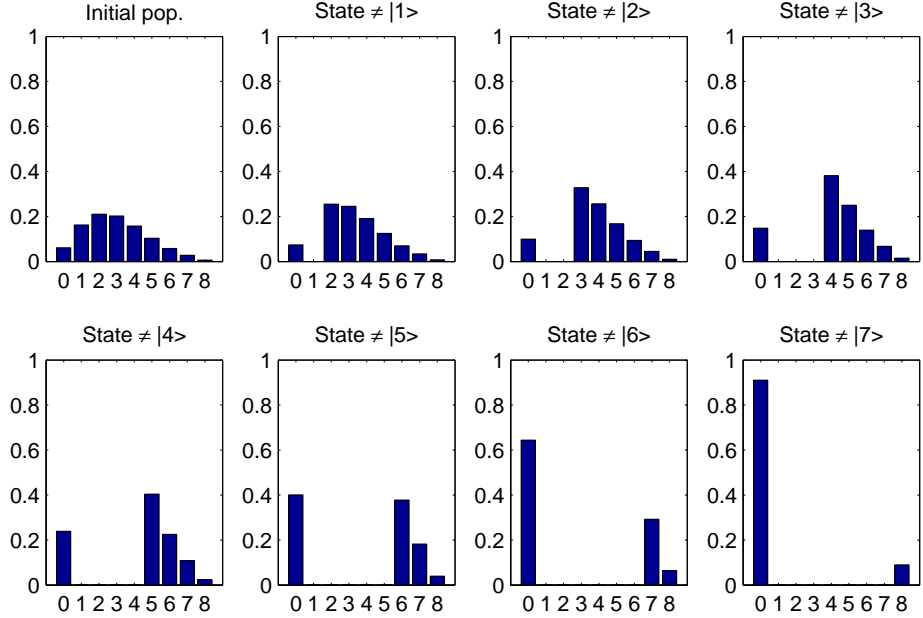}
\caption[justification = justified]{Evolution of the renormalized molecular rotational state population distribution after projecting the system to a new mixed state via heralded projection cooling, assuming $\alpha=1.0$}
\label{fig:population}
\end{figure}

\section{Molcular rotational sideband transitions with microwave fields}\label{sec:LambDicke}
A key element of the QLS protocol described in the previous section is the mapping of the rotational state onto the common motional state of both ions via a sideband transition driven by the spectroscopy field (step (iii)). For the field to couple to the motional state, it must vary significantly over the spatial extent of the molecular ion. The degree to which this is fulfilled is quantified by the so-called Lamb-Dicke parameter (LDP), $\eta=2\pi \sigma_0/\lambda$, which is a measure of the ground state rms spread of the ion's wavepacket, $\sigma_0$, relative to the wavelength, $\lambda$, of the electromagnetic field driving the transition. For a microwave transition in a molecular ion, at a wavelength on the cm scale and a wavepacket size of $\sim10$~nm, an LDP of $\sim10^{-6}$ or less is typical, which is about four orders of magnitude less than what is typically used in QLS. 

The problem of achieving an effective, non-negligible, LDP on a long-wavelength transition was first addressed by Mintert and Wunderlich~\cite{Mintert2001} and later by Ospelkaus et al.~\cite{Ospelkaus2008}. Both works considered magnetic field transitions between hyperfine levels of atomic ions and showed how an effective LDP could be tuned via a spatial gradient in a magnetic field. Recently, both of these schemes were implemented experimentally~\cite{Ospelkaus2011,Timoney2011}. 

Unfortunately, neither scheme extends straightforwardly to rotational state transitions in molecular ions. We stress that this is not true in general but it will be the case for a majority of diatomic hetero-nuclear molecules. Such molecules typically have as their electronic ground state the singlet $^1\Sigma$ state, which has zero angular momentum and the magnetic moment is therefore made up solely of the rotational magnetic moment and possibly a nuclear magnetic moment~\cite{Townes1955}. The rotational magnetic moment is generally very small, on the order of the nuclear magnetic moment, which means that the Zeeman shift of the rotational states will be about three orders of magnitude lower than for a state with non-zero angular momentum, such as the triplet $^3\Sigma$ state. A scheme relying on magnetic field interactions, as in Refs.~\cite{Mintert2001,Ospelkaus2008}, will thus require very steep and technically challenging magnetic field gradients. 

An alternative approach is to take advantage of the large dipole moment of polar molecules and couple their rotational states to an AC electric field gradient. Since this interaction can be facilitated with a laser, the experimental complexity is significantly reduced as this can be introduced in the experiment independently of the ion trap and vacuum chamber. 

In the following we first describe how to achieve an effective LDP via a spatial gradient field that couples to the states involved in the spectroscopic transition. We then show how this can be achieved in polar molecular ions via AC Stark shifts, induced by a far off-resonant laser beam. 

\subsection{Effective Lamb-Dicke parameter}
In the following we describe the basic principle of how an effective LDP emerges in a system, where the states involved in the relevant transition interact with a spatially varying external field. We note that this was already treated in Ref.~\cite{Mintert2001} but we detail the derivation here for completeness.

We consider a two level system of mass $m$ with states $|0\rangle$ and $|1\rangle$ in a harmonic potential with frequency $\omega_t$ along one motional eigen mode, which we define as the $z$-axis. These states could, e.g., be the $|0\rangle_\mathrm{rot}$ and the $|1\rangle_\mathrm{rot}$ rotational states of a polar molecule. The Hamiltonian for this system is
\begin{equation}
H = \frac{1}{2}\hbar \omega(z) \sigma_z + \hbar \omega_t a^{\dagger}a, 
\end{equation}
where $\sigma_z = |1\rangle\langle 1| - |0\rangle\langle 0|$, and $a$ and $a^\dagger$ are the annihilation and creation operators of the motional states, respectively. The energy difference between the two rotational states has a spatial dependence acquired through the interaction with some external potential $V_\mathrm{ext}(z)$:
\begin{equation}
\omega(z) = \frac{1}{\hbar} \left[\epsilon_1(V_\mathrm{ext}(z)) - \epsilon_0(V_\mathrm{ext}(z))\right].
\end{equation}
Inserting into the Hamiltonian, we get
\begin{equation}
H = \frac{1}{2}\left[\epsilon_1(V_\mathrm{ext}(z)) - \epsilon_0(V_\mathrm{ext}(z))\right] \sigma_z + \hbar \omega_t a^{\dagger}a.\end{equation}

At this point we make the assumption that the ion makes only small excursions around the trap location $z=0$ and that the energy of the two states is linear in their interaction with the external potential. This justifies only keeping terms to first order in an expansion of the energies of the two states ($i=0,1$):
\begin{eqnarray}
\epsilon_i &=& \epsilon_i(V_\mathrm{ext}(0)) + \frac{\partial \epsilon_i(V_\mathrm{ext}(z))}{\partial z}\Big |_{z=0}\times z\nonumber\\
&=& \epsilon_i(V_\mathrm{ext}(0)) + \partial_z \epsilon_i \times \sigma_0 (a + a^{\dagger}),
\end{eqnarray}
where we have expressed the position operator through the annihilation and creation operators and where we have defined $\sigma_0 = \sqrt{\frac{\hbar}{2m\omega_t}}$ as the ground state wave packet size. Inserting into the expression for the Hamiltonian we get
\begin{eqnarray}
H =\frac{1}{2}\hbar \omega_0 \sigma_z + \hbar \omega_t a^{\dagger}a + \frac{1}{2}\hbar \omega_t \epsilon_c (a + a^{\dagger}) \sigma_z,
\end{eqnarray}
where 
\begin{equation}
\omega_0 = \frac{1}{\hbar}\left[\epsilon_1(V_\mathrm{ext}(0))-\epsilon_0(V_\mathrm{ext}(0))\right] 
\end{equation}
and 
\begin{equation}
\epsilon_c=\frac{\sigma_0}{\hbar \omega_t}\left[\partial_z \epsilon_1-\partial_z \epsilon_0\right]_{z=0} .
\end{equation}

The last term describes a coupling between the rotational states and the motion of the molecular ion, which corresponds to the energy levels shifting as the ion is moving. This is of course what one expects from a potential gradient that couples to the energy levels; nevertheless, it is useful for our treatment here to transform into a frame in which the energy levels in the presence of the external potential are constant. This is achieved through the unitary transformation $\tilde{H} = e^{iS} H e^{-iS}$, where $S = \frac{1}{2i}\epsilon_c (a^\dagger - a)\sigma_z$. Invoking this, the transformed Hamiltonian reads
\begin{equation}
\tilde{H} = \frac{1}{2}\hbar\omega_0\sigma_z + \hbar \omega_t a^\dagger a - \frac{1}{4}\hbar \omega_t \epsilon_c^2.
\end{equation}
The last constant term does not affect the dynamics and will therefore be ignored in the following. 

We now introduce an electromagnetic field that interacts with the molecular ion on, or near, the $|0\rangle_\mathrm{rot} \leftrightarrow |1\rangle_\mathrm{rot}$ transition. The corresponding Hamiltonian reads: 
\begin{equation}\label{eq:HamiltonE}
H_E = \frac{1}{2}\hbar\Omega (\sigma_+ + \sigma_-)(e^{i[k_z z - \omega_E t]} + \mathrm{H.c.}),
\end{equation}
where $\Omega=\frac{\bmu\cdot \mathbf{E}}{\hbar}$ is the Rabi frequency, with $\bmu$ being the relevant transition dipole moment and $E$ the electromagnetic field strength. $\omega_E$ is the frequency of the field and $k_z = \frac{\omega_E}{c}\cos \Theta$ is the wave vector, with $\Theta$ accounting for the angle between the incident beam and the $z$-axis. Note that we are using the subscript $E$ to signify that the transition of interest is an electric dipole transition, specifically, the $|0\rangle_\mathrm{rot} \leftrightarrow |1\rangle_\mathrm{rot}$ rotational transition. 

We replace the position coordinate by its quantum mechanical operator, $z = \sigma_0(a^\dagger + a)$, and rewrite \Eref{eq:HamiltonE} in terms of the LDP $\eta = \sigma_0 k$:
\begin{equation}H_E = \frac{1}{2}\hbar\Omega (\sigma_+ + \sigma_-)(e^{i[\eta (a^\dagger + a) - \omega_E t]} + \mathrm{H.c.}).
\end{equation}
The interaction Hamiltonian is transformed as above according to $\tilde{H}_E = e^{iS} H_E e^{-iS}$, which gives
\begin{eqnarray}
\tilde{H}_E 
&=& \frac{1}{2}\hbar\Omega (\sigma_+ e^{\epsilon_c (a^\dagger - a)} + \sigma_- e^{-\epsilon_c (a^\dagger - a)})\nonumber\\
&\times&(e^{i[\eta (a^\dagger + a -\epsilon_c\sigma_z) - \omega_E t]} + \mathrm{H.c.})
\end{eqnarray}

It is useful to transform $\tilde{H}_E$ into the interaction picture with respect to $\tilde{H}$, i.e., performing the transformation $\tilde{H}_E^\mathrm{int} = e^{i\tilde{H} t/\hbar}\tilde{H}_Ee^{-i\tilde{H} t/\hbar}$, where $\tilde{H} = \frac{1}{2}\hbar\omega_0\sigma_z + \hbar \omega_t a^\dagger a$ as found above. Invoking the rotating wave approximation, omitting terms oscillating at $2\omega_0$, the final result for the interaction Hamiltonian reads:
\begin{eqnarray}\tilde{H}_E^\mathrm{int}&=&\frac{1}{2}\hbar\Omega \Big(\sigma_+ e^{-i(\Delta t - \eta\epsilon_c)} \times e^{i[(\eta + i\epsilon_c)a + (\eta - i\epsilon_c)a^\dagger]}\nonumber\\
&+& \sigma_- e^{i(\Delta t + \eta\epsilon_c)}\times e^{-i[(\eta + i\epsilon_c)a + (\eta - i\epsilon_c)a^\dagger]}\Big),
\end{eqnarray}
where $\Delta = \omega_E - \omega_0$ is the frequency detuning.

From the above expression for the interaction Hamiltonian, we can define an effective, complex LDP: $\eta_\mathrm{eff} = \eta + i\epsilon_c $. Rewriting this complex parameter in terms of its magnitude and phase, we get $\eta_\mathrm{eff} = \sqrt{\eta^2 + \epsilon_c^2} e^{i\phi}$, where the phase $\phi = \arctan (\epsilon_c/\eta)$ for $\eta > 0$. The interaction Hamiltonian can now be written as:
\begin{eqnarray}
\tilde{H}_E^\mathrm{int}= \frac{1}{2}\hbar\Omega\left(\sigma_+ e^{-i\Delta t} \times e^{i\sqrt{\eta^2 + \epsilon_c^2}[ae^{i\phi} +a^\dagger e^{-i\phi}]} + H.c.\right),\nonumber
\end{eqnarray}
where we have absorbed the $e^{i\eta\epsilon_c}$ phase into $\sigma_+$, which is only defined up to an arbitrary phase anyway. Similarly, we may absorb the $e^{\pm i\phi}$ phases into the $a$ and $a^\dagger$, in which case the interaction Hamiltonian is expressed simply as
\begin{equation}
\tilde{H}_E^\mathrm{int} = \frac{1}{2}\hbar\Omega\left(\sigma_+ e^{-i\Delta t} \times e^{i|\eta_\mathrm{eff}|[a +a^\dagger]} + H.c.\right).
\end{equation}
The key result of this derivation, as was first pointed out by Mintert and Wunderlich~\cite{Mintert2001}, is that an effective LDP, $|\eta_\mathrm{eff} |= \sqrt{\eta^2 + \epsilon_c^2}$, has emerged, and that this is tunable through $\epsilon_c$. This means that even when $\eta\simeq 0$, as is typically the case on a microwave transition owing to the long wavelength, we have $|\eta_\mathrm{eff}|\approx \epsilon_c$. Sideband transitions are still feasible when $\epsilon_c$ takes on non-negligible values. In the following, we shall consider a method for achieving non-zero values of $\epsilon_c$.

\subsection{Optical AC Stark shift gradient}
A large effective LDP can be achieved using a spatially varying laser field, provided the dynamic polarizabilities at the laser frequency of the two rotational states are sufficiently different. Calculations as well as data for the dynamic polarizability of molecular ions are scarce in the litterature. Neutral molecules, however, have been studied in much greater detail and in particular the alkali dimers such as KRb and RbCs~\cite{Kotochigova2006,Kotochigova2010}. Experiments with these molecules typically seek to achieve confinement in optical dipole traps using far off-resonant lasers, which may induce significant AC Stark shifts, also on the rotational states. For this reason, there have been strong efforts to find so-called \emph{magic} wavelengths of the trapping laser, where the differential Stark shift between the $|0\rangle_\mathrm{rot}$ and the $|1\rangle_\mathrm{rot}$ states is zero. While our goal here is precisely the opposite, we may still rely on the quantitative analysis from this work in evaluating the feasibility of our technique.

For the purpose of providing an order of magnitude estimate of the electric field gradient needed to achieve our proposed scheme, we consider the $\mathrm{Sr}\mathrm{Cl}^+$ molecular ion, which is similar in size and mass as KRb for which detailed studies of the dynamic polarizability exist. We further justify this simplification in our estimate by noting that, despite the difference in atomic species, the values obtained for KRb and RbCs are of the same order of magnitude~\cite{Kotochigova2006,Kotochigova2010}. By the same reasoning, we would expect similar behavior for the $\mathrm{Sr}\mathrm{Cl}^+$ molecular ion.

Based on the value for the difference in polarizability of the $|0\rangle_\mathrm{rot}$ and $|1\rangle_\mathrm{rot}$ states, $\Delta \alpha_{01}$, for KRb reported in, e.g., Ref.~\cite{Kotochigova2010}, we give an estimate for $\mathrm{Sr}\mathrm{Cl}^+$ of $\Delta\alpha_{01}\sim 1000$ a.u., where 1 a.u.~$\sim 4.7\times10^{-8}$~MHz/(W/cm$^2$). We note that the calculations of Ref.~\cite{Kotochigova2010} were performed for a wide spectrum of wavelengths from the visible and into the IR and that $\Delta \alpha_{01}$ is expected to be fairly constant across this range apart from regions around electronic resonances. This suggests that one can rely on wavelengths around $1~\mu$m or $1.5~\mu$m where high-power, stable sources are commercially available.

The change in energy due to the AC Stark shift is given by $\Delta E(z) = \hbar \Delta \alpha_{01} I(z)$. We consider both a standing wave laser field along the $z$-axis as well as a traveling wave laser field propagating perpendicular to the $z$-axis and focused to a tight waist, $w_0$. In the former case $I(z)=I_0 \sin^2(kz+\phi)$, while in the latter $I(z)=I_0 e^{-(z-z_0)^2/w_0^2}$, where in both cases the peak intensity is $I_0=\frac{2P}{\pi w_0^2}$, with the $P$ being the total power in the beam. In both cases we assume that the molecular ion is positioned at the point of steepest optical gradient, corresponding to $\phi=\frac{\pi}{4}$ and $z_0=w_0$ in the standing and traveling wave cases, respectively. We furthermore assume that the wavepacket size is small compared to the optical wavelength (Lamb-Dicke regime), which motivates an expansion around $z=0$ to get:
\begin{equation}\label{Eq:Eshift_effLDparam}
\Delta E(z) = 
\cases{
\hbar \Delta \alpha_{01} I_0 \left(\frac{1}{2}+kz\right), \quad\textmd{standing wave,} \\ 
\hbar \Delta \alpha_{01} I_0 e^{-1} \left(1-\frac{2z}{w_0}\right), \quad\textmd{traveling wave.}
}
\end{equation}
Using $z=\sigma_0 (a+ a^\dagger)$, the expression for the effective LDP can be found to be
\begin{equation} \label{Eq:effLDparam}
|\eta_\mathrm{eff}|\approx\epsilon_c = 
\cases{
\frac{\Delta \alpha_{01} }{\omega_t}k\sigma_0\times I_0,\quad \textmd{standing wave,} \\ 
\frac{\Delta\alpha_{01}}{\omega_t} e^{-1} \frac{2\sigma_0}{w_0}\times I_0, \quad\textmd{traveling wave.}
}
\end{equation}

\begin{figure}
\centering
\includegraphics[trim = 28mm 0mm 30mm 0mm, clip, width=1.0\columnwidth]{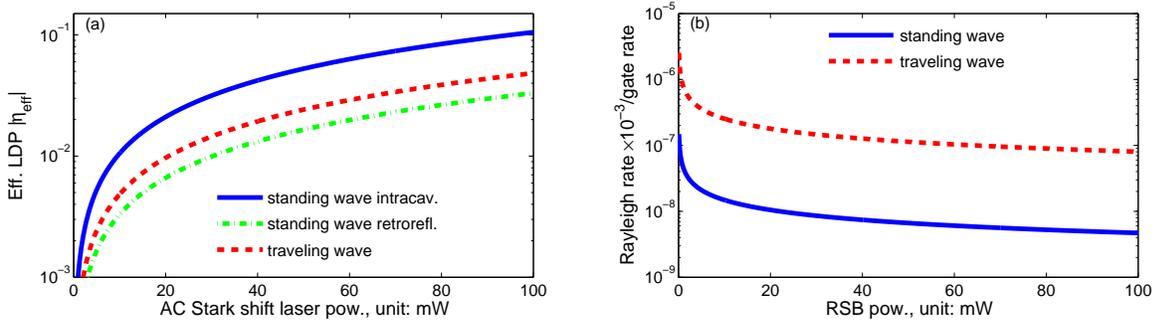}
\label{fig:EffLDPscatt}
\caption{(a) Effective LDP produced by spatially varying laser field for three different scenarios, all at wavelength $\lambda=1~\mu$m. Red: focused laser beam, $w_0=2~\mu$m. Green: retro-reflected laser beam, $w_0=20~\mu$m. Blue: Cavity laser beam, $w_0=100~\mu$m, cavity finesse, $F=1000$. (b) Estimate of spontaneous-Raman-rate-to-Rabi-frequency ratio. Raman scattering rate is estimated using $10^{-3} \times$ Rayleigh rate, for travelling and standing waves, both at wavelength $\lambda = 1~\mu$m.}
\end{figure}

The standing wave configuration can be realized either by retro-reflecting the laser beam off a single mirror or by using an optical cavity. Figure~\ref{fig:EffLDPscatt} (a) plots the effective LDP for three different experimental scenarios, namely, a tightly focused laser beam, a retro-reflected, moderately tight focused laser beam, and a loosely focused cavity beam, all with wavelength $\lambda=1~\mu$m. For each scenario the focusing is chosen to allow for an effective LDP in the $10^{-2}-10^{-1}$ range with reasonable power level and the figure thus represents the compromise between complexity and power available that an experiment will be subject to. The example is based on $\mathrm{SrCl^+}$ confined in a trap with a trap frequency, $\omega_t/2\pi=1$~MHz and $\Delta\alpha_{01}\sim1000$~a.u..

It is clear that an appreciable LDP can be achieved. Even larger values of $\epsilon_c$ could be achieved without going to higher power levels by lowering the trap frequency \cite{Poulsen2012arXiv}, which would increase $\sigma_0$ accordingly. Ultimately, the trap frequency is expected to be constrained by the requirement of the atomic ion remaining in the Lamb-Dicke regime in order to obtain efficient sideband operations as part of steps (iii), (iv) and (v) of Section~\ref{sec:QLS}. Recently, however, sideband cooling to the ground state was demonstrated at trap frequencies lower than 300~kHz~\cite{Poulsen2012PRA}.

\subsection{Decoherence induced by off-resonance laser field}
We have shown that an effective LDP can be achieved by coupling the trapped molecule with an off-resonant laser field. However, the increase in effective LDP with the laser intensity, as captured by Figure~\ref{fig:EffLDPscatt} (a), is accompanied by an increase in the scattering rate of the molecule. In this respect, the spontaneous Raman scattering rate is of particular relevance as it can lead to redistribution of population to different rovibrational levels. The Raman scattering rate can be gauged from the Rayleigh scattering rate, which can be estimated simply from knowledge of the polarizability, $\alpha$, alone~\cite{Tannoudji1992a}:
\begin{equation}
R = \frac{2\pi \hbar \omega_E^3 \alpha^2}{3 c^2}\times I_0.
\end{equation}
The linear dependance on intensity, $I_0$, similarly to the expression for $\epsilon_c$ (\Eref{Eq:effLDparam}), suggests that the optimal intensity in a given experiment will be determined by the microwave Rabi frequency and the coherence time of the rotational states. The spontaneous Raman scattering rate is typically about three orders of magnitude lower than the Rayleigh scattering rate~\cite{Long1977}, which means that an experiment will, in general, have to fulfill the inequality:
\begin{equation}
\label{eq:gateVSdec}
\eta_\mathrm{eff}\Omega_\mathrm{J,J'}\gg R\times10^{-3}, \gamma_\mathrm{J,J'}, \dot{n},
\end{equation}
where $\Omega_\mathrm{J,J'}$ is the Rabi frequency on the $J$ to $J'$ rotational microwave transition, $\gamma_\mathrm{J,J'}$ is the decoherence rate and $\dot{n}$ is the heating rate of the common motional mode of the ion and molecule resulting from their coupling to the environment~\cite{Turchette2000}. 

The microwave Rabi frequency can be evaluated from the electric field strength and the dipole moment. For instance, an electric field on the order of 5~V/m can be straightforwardly produced from a microwave horn located outside the vacuum chamber, and a dipole moment on the order of the 10~Debye is common to polar molecular ions such as SrCl$^+$ and CaCl$^+$. Under these conditions, a sideband Rabi frequency on the order of $2\pi\times$100~kHz can be achieved. A typical value for $\gamma_\mathrm{J,J'}$ is on the order of 10~Hz \cite{Schuster2011}. Heating rate has been measured to be a few Hz at a cryogenic temperature \cite{Labaziewicz2008}, and we take a conservative estimate of $\sim$1 kHz \cite{Seidelin2006}. The ratio $\frac{R\times10^{-3}}{\eta_\mathrm{eff}\Omega_\mathrm{J,,J'}}$ is plotted in Figure~\ref{fig:EffLDPscatt} (b). We see that at reasonable power for the microwave addressing the molecular transitions, the ratio of Raman dechoherence rate and the Rabi frequency can be made as low as $10^{-5}$. The condition as given in \Eref{eq:gateVSdec} is clearly satisfied, showing a reasonable feasibility for our experimental protocol. 

\section{Quantum information processing with polar molecular ions}\label{sec:QIPpolmol}

Polar molecular ions are attractive candidates for quantum information processing owing to the long coherence time and relatively large dipole moment of their rotational states. In a quantum processor, these characteristics lead to a favorable ratio between the rate at which information can be manipulated and the rate at which information is lost. For a quantum computer to perform successfully, the former should obviously exceed the latter and, preferably, by several orders of magnitude~\cite{Steane2007}. As we showed above, Rabi-frequencies for rotational transitions of order $\sim 100$~kHz are realistic in the systems considered here. This greatly exceeds the decay rate of the rotational transitions by 4 orders of magnitude and illustrates the potential of the system as a quantum information processor if rotational transitions are used as qubit states. A further highly attractive technical benefit to the use of molecular rotational states is that they can be manipulated using microwave fields for which reliable and stable sources are readily available. This significantly reduces the experimental overhead compared with implementations based on trapped atomic ions, which typically requires highly-stable laser sources in the visible and even uv-range~\cite{Blatt2008}. 

Hitherto, there have been four significant challenges opposing the implementation of a quantum information processor based on polar molecules and coherent microwave fields: (i) The long wavelengths of the microwave transitions implies that there is negligible coupling to the motion of the molecular ion and, hence, any conditional multi-qubit gate, such as that of Cirac and Zoller~\cite{Cirac1995} or M\o lmer and S\o rensen~\cite{Molmer1999}, that relies upon a motional bus, will be very slow. (ii) Individual addressing of molecular ions is also complicated by the use of long ($\sim$cm) wavelengths due the typical ion spacings in ion traps being on of the order of $10~\mu$m. (iii) The absence of closed transitions between the rovibrational states means that conventional laser cooling techniques fail in preparing the molecule in its rotational ground state. (iv) Finally, qubit state detection is similarly complicated by the lack of a cycling transition. 

We have already outlined methods to overcome challenges (i), (iii) and (iv) in Section~\ref{sec:LambDicke}, ~\ref{sec:GroundStateCooling} and \ref{sec:QLS}, respectively. As for individual addressing, since the ions are typically only $\sim 10$~$\mu$m apart, the microwave fields will in general address more than a single ion. However, this issue is also resolved by the strong laser field used to generate the effective LDP. As seen from \Eref{Eq:Eshift_effLDparam}, there is a constant shift of $\Delta E\sim\hbar \Delta \alpha_{01} I_0$, which for the parameters used above, is on the order of 1~MHz. The natural line width of the rotational transition is extremely narrow and as discussed in Ref.~\cite{Schuster2011}, even in the presence of stray DC fields and other imperfections, values in the range of a few tens of Hz are realistic. This implies that the AC Stark shift will significantly isolate the laser addressed molecular ions in frequency space from non-addressed molecular ions. Furthermore, the cotrapped atomic-molecular-ion system can be used as a hybrid quantum information processor, where fast gate operations with atomic ions and long-lived quantum memories with moleculalr ions can both be exploited.

\section{Conclusions and discussion}\label{sec:Conclusions}

In summary, we propose a novel approach to study the rotational spectroscopy of trapped polar molecular ions, in the framework of quantum logic spectroscopy. We show that direct microwave addressing can be achieved via an effective Lamb Dicke parameter on the order of $\sim$0.1, induced by a spatially varying AC Stark shift, which significantly reduces the experimental complexity. We show that under our particular QLS protocol, all cooling on the rotational state of the molecule can be achieved using heralded projection.

The protocol of projected cooling, state detection, and induced effective Lamb Dicke parameter has shown great promise to implement polar molecular ion quantum information processing. Possible implementation of individual qubit addressing in a molecular ion chain can also be readily achieved with an AC Stark-shift-inducing laser, albeit the long wavelength of microwaves.

\ack

M.~S.~Thanks Michael Gutierrez, Alexei Bylinskii, Tout Wang, and Wujie Huang for helpful discussions. This work was supported by the NSF Center for Ultracold Atoms, the DARPA QUEST program, and the AFOSR MURI on ultracold molecules. P.~F.~H. is grateful for the support from the Carlsberg Foundation and the Lundbeck Foundation. M.~D. appreciates support through the Danish National Research Foundation Center for Quantum Optics - QUANTOP, the The Danish Agency
for Science, Technology and Innovation, the Carlsberg
Foundation as well as the Lundbeck Foundation.


\section*{References}

\bibliography{bibliographyNJPnoURL0701.bib}

\end{document}